# Photochemical Haze Formation in the Atmospheres of super-Earths and mini-Neptunes

Chao He[1], Sarah M. Hörst[1], Nikole K. Lewis[1,2], Xinting Yu[1], Julianne I. Moses[3], Eliza M.-R. Kempton[4], Mark S. Marley[5], Patricia McGuiggan[6], Caroline V. Morley[7], Jeff A. Valenti[2], & Véronique Vuitton[8]

[1] Department of Earth and Planetary Sciences, Johns Hopkins University, Baltimore, MD, USA che13@jhu.edu

[2]Space Telescope Science Institute, Baltimore, MD, USA

[3]Space Science Institute, Boulder, CO, USA

[4]Grinnell College, Grinnell, IA, USA

[5]NASA Ames Research Center, Mountain View, CA, USA

[6]Department of Materials Science and Engineering, Johns Hopkins University, Baltimore, MD, USA

[7]Harvard University, Cambridge, MA, USA

[8]Université Grenoble Alpes, Grenoble, FR

**Abstract:**

UV radiation can induce photochemical processes in exoplanet atmospheres and produce haze particles. Recent observations suggest that haze and/or cloud layers could be present in the upper atmospheres of exoplanets. Haze particles play an important role in planetary atmospheres and may provide a source of organic material to the surface which may impact the origin or evolution of life. However, very little information is known about photochemical processes in cool, high-metallicity exoplanetary atmospheres. Previously, we investigated haze formation and particle size distribution in laboratory atmosphere simulation experiments using AC plasma as the energy source. Here, we use UV photons to initiate the chemistry rather than the AC plasma, since photochemistry driven by UV radiation is important for understanding exoplanet atmospheres. We present photochemical haze formation in current UV experiments; we investigated a range of atmospheric metallicities (100×, 1000×, and 10000× solar metallicity) at three temperatures (300 K, 400 K, and 600 K). We find that photochemical hazes are generated in all simulated atmospheres with temperature-dependent production rates: the particles produced in each metallicity group decrease as the temperature increases. The images taken with atomic force microscopy show the particle size (15-190 nm) varies with temperature and metallicity. Our laboratory experimental results provide new insight into the formation and properties of photochemical haze, which could guide exoplanet atmosphere modeling and help to analyze and interpret current and future observations of exoplanets.

## 1. INTRODUCTION

Over 3500 exoplanets have been confirmed and the majority of them are super-Earths and mini-Neptunes (generally any planet with size or mass between Earth's and Neptune's) (e.g., Borucki et al. 2011, Fressin et al. 2013). These planets are likely to have different kinds of compositions in their atmospheres (e.g., Elkins-Tanton & Seager 2008, Miller-Ricci et al. 2009, Schaefer et al. 2012, Moses et al. 2013, Hu & Seager 2014, Venot et al. 2015, Ito et al. 2015). The Transiting Exoplanet Survey Satellite (TESS) mission, successfully launched in April 2018, will find more super-Earths and mini-Neptunes, whose atmospheres will be characterized by future telescopes, including the James Webb Space Telescope (JWST). The atmospheres of a number of small planets ($R_p$ < $R_{Neptune}$) with cool temperatures ($T_{eq}$ <1000 K) have now been observed, and a majority of these planets show evidence for aerosols (clouds or hazes) (e.g., Kreidberg et al. 2014; Knutson et al. 2014a, 2014b; Dragomir et al. 2015). Condensate cloud and photochemical haze particles are present in many solar system atmospheres. They are also expected in exoplanet atmospheres based on our understanding of particle formation in planetary atmospheres.

Particles play an important role in planetary atmospheres and can affect atmospheric chemistry and thermal profile. Photochemically generated hazes may provide a source of organic material to the surface which may impact the origin or evolution of life. The photochemistry induced by UV photons is universal in planet systems. Solar UV photons drive the photochemistry in atmospheres of solar system bodies (such as Venus, Earth, Jupiter, Saturn, Titan, Triton, and Pluto), and stellar UV radiation also induces photochemistry in the atmospheres of exoplanets. Studies show that the UV radiation around M dwarf planet hosts covers both far UV and near UV wavelengths; thus inclusion of UV driven atmospheric chemistry is important for understanding atmospheres of planets orbiting most M dwarfs (e.g., France et al. 2013). Photochemistry is likely to play a major part in the atmospheres of planets with $T_{eq}$ <1000 K, especially for super-Earths and mini-Neptunes that may have enhanced atmospheric metallicity (Marley et al. 2013). Metallicity (Z) is defined as the fractional percentage of the chemical elements other than hydrogen and helium in a star or other object (solar metallicity, $Z_{sun}$=0.0134). However, we currently have very little information about photochemical processes in these cool, metal-rich planetary atmospheres. Laboratory production and analysis of exoplanet hazes are essential for interpreting future spectroscopic observations and properly characterizing the atmospheres of these worlds. Recently, we conducted laboratory atmosphere simulation experiments with AC plasma as energy source and reported the haze formation and particle size distribution in these experiments (He et al. 2018, Hörst et al. 2018a) that explored a broad range of atmospheric phase space relevant to super-Earths and mini-Neptunes. Here, we continue our investigation in the same parameter space by using a different energy source, UV photons. We show that photochemical hazes are generated in these diverse atmospheres, and the haze production rate and the particle size varies with temperature and metallicity.

## 2. MATERIALS AND EXPERIMENTAL METHODS

### *2.1. Haze Production Setup*

We performed the UV experiments by using the Planetary Haze Research (PHAZER) experimental setup at Johns Hopkins University (He et al. 2017) with the same gas mixtures as our previous plasma experiments (He et al. 2018, Hörst et al. 2018a). A schematic of the setup and the initial gases are shown in Figure 1. As we discussed in previously (He et al. 2018, Hörst et al. 2018a), chemical equilibrium models were used for gas mixture calculations (Moses et al. 2013). The calculated gas mixtures are a good starting point to investigate the photochemical processes in the atmospheres of super-Earths and mini-Neptunes with higher metallicity (100×, 1000×, and 10000× solar metallicity) at 300, 400, and 600 K. More details of the gas mixtures and the experimental procedure can be found in He et al. (2018). The prepared gas mixture is heated to desired temperature (600 K, 400 K, or 300 K), and flows through the reaction chamber. A hydrogen UV lamp (HHeLM-L, Resonance LTD.) is attached to the chamber, producing UV photons that initiate photochemical processes in the gas mixture. The lamp was set to produce continuum UV photons from 110 nm to 400 nm. The total UV flux of the lamp is about $3\times10^{15}$ photons/(sr*s), and the VUV and UV output spectrum can be found at www.resonance.on.ca. Lamps with similar wavelength range and flux are used for simulating photochemistry in atmosphere of early Earth and Titan (see e.g., Trainer et al. 2006, 2012; Sebree et al. 2014; Hörst et al. 2018b). Although the photons in this wavelength range could not dissociate molecules like $N_2$ or CO directly, previous studies show that incorporation of nitrogen in organic products produced from $N_2/CH_4$ mixtures that were irradiated with similar UV lamps, suggesting an unknown photochemical process is occurring to incorporate N into the molecular structure of the aerosol (Hodyss et al. 2011, Trainer et al. 2012). The organics produced in these experiments could be the source for life to arise (e.g., Miller 1953; Sagan & Khare 1971; Trainer et al. 2006, 2012; Hörst et al. 2012, 2018b), since many nitrogenous molecules, such as amino acids and nucleobases, are building blocks of life.

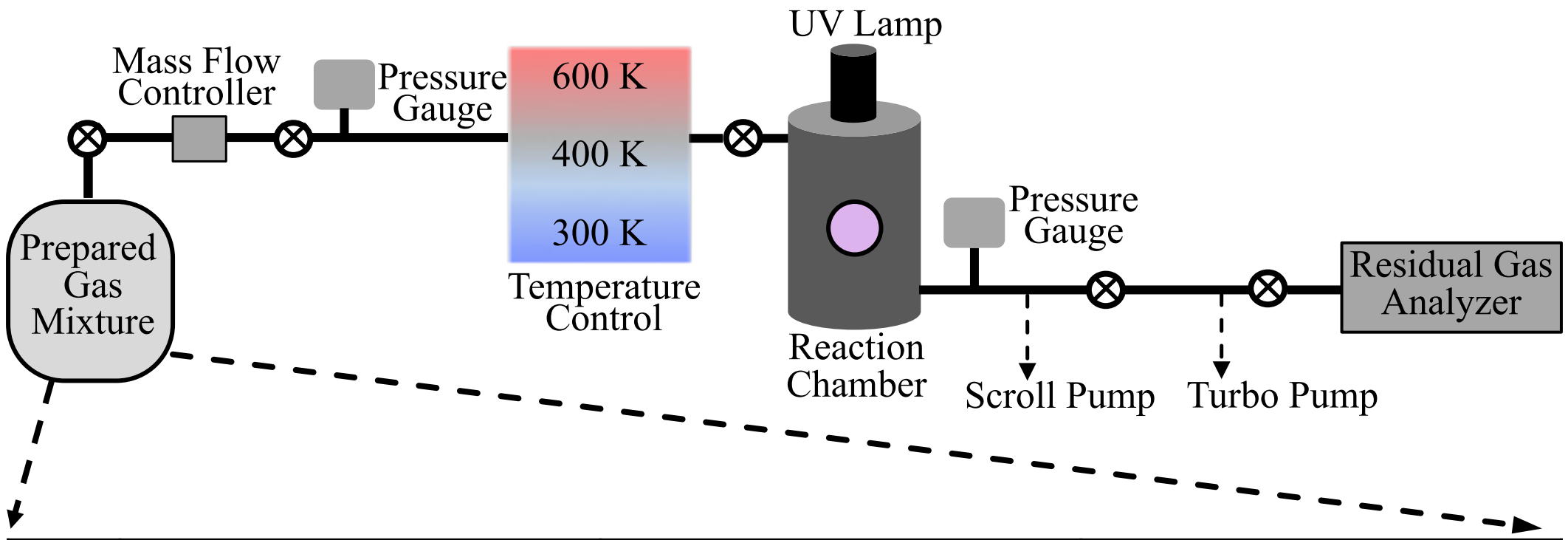

| | 100× Solar Metallicity | 1000× Solar Metallicity | 10000× Solar Metallicity |
|---|---|---|---|
| 600 K | 18.3% He, 72.0% $H_2$, 6.3% $H_2O$, 3.4% $CH_4$ | 13.3% He, 42.0% $H_2$, 20.0% $CO_2$, 16% $H_2O$, 5.1% $N_2$, 1.9% CO, 1.7% $CH_4$ | 4.1% He, 66.0% $CO_2$, 12.0% $N_2$, 8.6% $H_2$, 5.9% $H_2O$, 3.4% CO |
| 400 K | 17.2% He, 70.0% $H_2$, 8.3% $H_2O$, 4.5% $CH_4$ | 14.7% He, 56.0% $H_2O$, 11.0% $CH_4$, 10.0% $CO_2$, 6.4% $N_2$, 1.9% $H_2$ | 5.0% He, 67.0 % $CO_2$, 15.0% $H_2O$, 13.0% $N_2$ |
| 300 K | 17.3% He, 68.6% $H_2$, 8.4% $H_2O$, 4.5% $CH_4$, 1.2% $NH_3$ | 16.0% He, 66.0% $H_2O$, 6.6% $CH_4$, 6.5% $N_2$, 4.9% $CO_2$ | 4.1% He, 67.3 % $CO_2$, 15.6% $H_2O$, 13.0% $N_2$ |

Figure 1. Schematic of the PHAZER setup for current work and the gas mixtures used in the experiments. To be noted, the schematic is shown to provide a concept of our setup, and it is updated from our previous studies to include the new energy source (UV lamp); the gas mixtures used here are the same as our previous plasma experiments (He et al. 2018, Hörst et al. 2018a). The detailed procedure was discussed in He et al. (2018).

The gases flow continuously under UV irradiation for 72 hr and solid particles (if produced in the experiment) are deposited on the chamber's inside wall and quartz substrates (Ted Pella, Inc., the same substrates as used in our previous plasma experiments). We ran our AC glow discharge (plasma) experiments for 72 hr under the same conditions (He et al. 2018, Hörst et al. 2018a), thus we also ran the UV experiments for the same amount of time for comparison. We collected the samples in a dry $N_2$ glove box, and followed the same collection procedure as our previous studies (He et al. 2018, Hörst et al. 2018a).

### *2.2. Atomic Force Microscopy (AFM) Measurement*

Several different techniques have been used to measure the particle sizes of Titan haze analogs prepared in laboratories, which were summarized in He et al. (2018). As discussed in our previous study (He et al. 2018), AFM was chosen in our study since it is a non-destructive technique and does not require special sample preparation. An atomic force microscope (Bruker Dimension 3100, Bruker Nano) was used to examine the surface morphology of the particles on the quartz substrates. The tip (silicon probe, Tap300-G, Ted Pella, Inc) and the setting (tapping mode) for the measurement are the same as our previous study (He et al. 2018).

## 3. RESULTS AND DISCUSSION

### *3.1. Photochemical Haze Formation and AFM Images of the Particles*

In our plasma experiments, all simulated atmospheres produced particles, but the particle production rate varied substantially, as high as 10 mg/hr for the cooler (300 and 400 K) 1000× metallicity experiments (Hörst et al. 2018a). For our current UV experiments, no particles were observed by visual inspection on the walls of the chamber after 72 hr flow under UV irradiation. No obvious difference between the quartz disks from the experiments and the blank quartz disk (not exposed to UV or experimental gas mixtures) could be visually observed. This suggests that the production rates must be very low even if the photochemical processes generate haze particles. Compared to the clear disks from current UV experiments, our previous plasma experiments for the same gas mixtures produced many more particles and formed colorful films (He et al. 2018). It is not surprising since the haze production rates from UV experiments are usually lower than those from plasma experiments (Peng et al. 2013, Hörst et al. 2018b).

Since the particle production rates are relatively low in the UV experiments, it is hard to tell whether or not haze particles are produced in these experiments by visual examination. Thus, we observed the disks under AFM. Figure 2 shows AFM images of these disks, a clean blank quartz disk and a disk from the 600 K reference experiment, displaying 1 μm x 1 μm scanning area for each one. AFM image displays that the blank disk has a smooth, clean surface. The AFM images show that the disk surface from the reference experiment is also smooth and clean, indicating that there are no particles produced from the gases without UV exposure. Compared to the blank and reference disks, spherical particles are observed on the disks from all nine experiments, indicating that haze particles are produced from photochemical processes in all nine diverse atmospheres. Figure 2 shows that the number and size of the particles from these experiments have great variations with the different gas mixtures at different temperature. There are numerous small particles produced from the 400 K experiments, while the 600K and 300 K experiments generate fewer particles with broader size range. For all compositions, the 400 K experiments produced the smallest particles and the 300 K experiments typically produced the largest particles, despite the fact that the initial gas compositions were very similar at these two temperatures. The 600 K experiments had particle sizes intermediate to those at 300 and 400 K. We can determine the particle diameter from the AFM images by measuring the projection of the particle on the x-y plane. The diameter measured from this method is relatively accurate in the particle size range from our experiments (diameter errors < 3 nm), as we discussed in previous study (He et al. 2018). For the 600 K experiments, the particles (diameter 20 to 110 nm) are sparsely spread on the disks, while for the 400 K experiments, many small particles (15 to 60 nm) are closely compacted on the substrate. The 300 K cases produce particles with wider size range (diameter 35 to 190 nm).

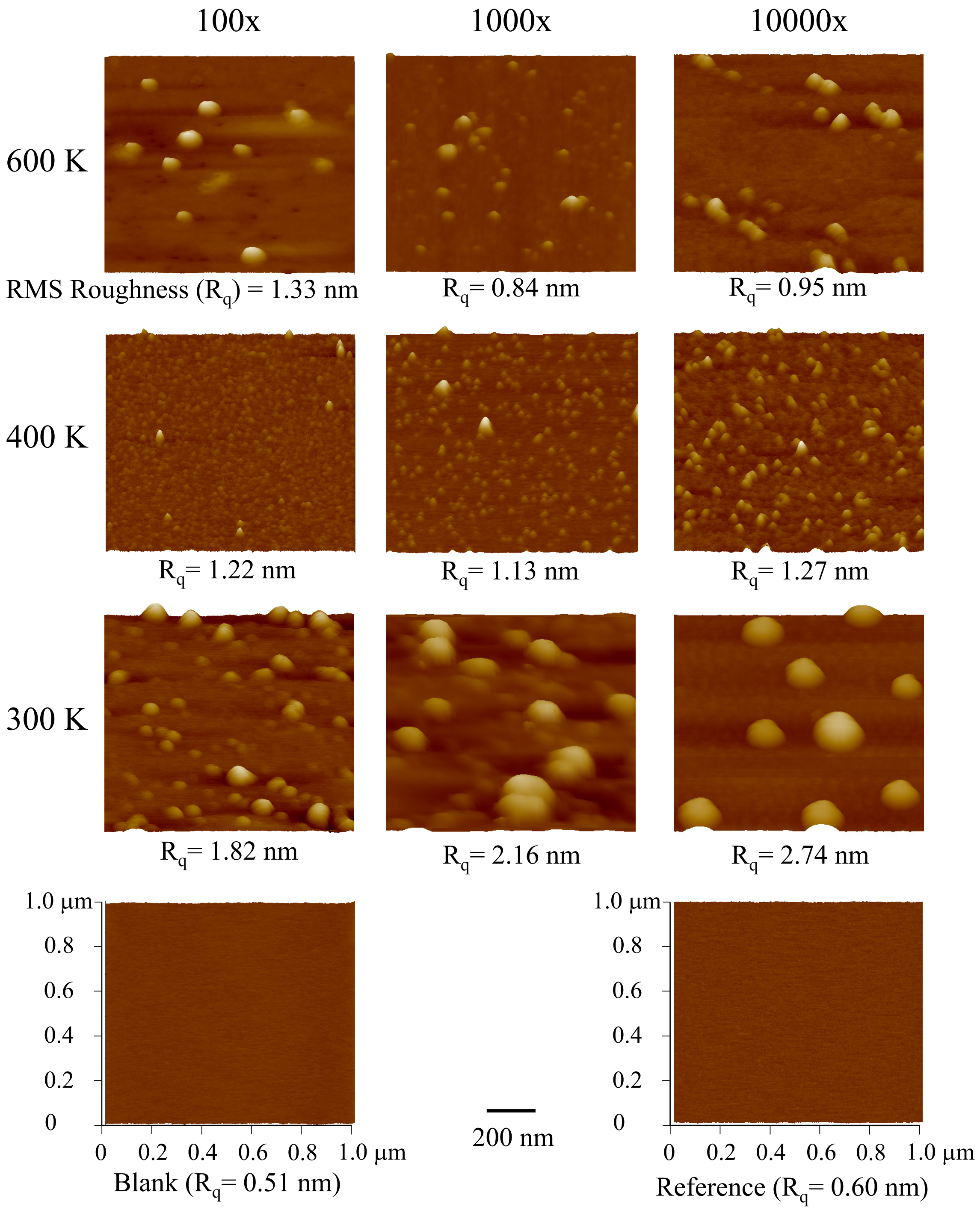

Figure 2. AFM 3D images of the particles on quartz substrates. Scanning area is 1 µm×1 µm for each sample. Blank is the image from a clean blank quartz disk. Reference is the AFM image of the disk from the reference experiment without UV exposure at 600 K for the 10000× metallicity case. The height scale of the AFM images is 50 nm for the 300 K and 600 K experiments, 20 nm for the 400 K experiments to better show the small particles, and also 20 nm for the blank disk.

The small number of the root mean square roughness ($R_q$) indicates these films are very smooth.

As shown in the AFM images (Figure 2), the haze particles are produced from photochemical processes in all nine diverse atmospheres. However, the production rates are so low in these UV experiments, and there is not enough solid produced to collect and weigh. Therefore, it is difficult to compare the production rates of these experiments. Due to the low production rate, the particles are assumed to be deposited as single layer on the quartz disk, as shown in Figure 2. Very few or no aggregates are observed in the AFM images, indicating that most of the haze particles produced in the phase space we investigated here are monomers. The roughness (< 3 nm) indicates the films are smooth, supporting single layer of monomers on the disks. The general size range from all nine experiments is from 15 nm to 190 nm, which is similar to that (20 nm to 180 nm) from our previous plasma experiments at the same conditions. This is a relatively narrow range, considering the huge differences in the gas compositions, temperatures, and energy sources. It could imply some similarity in the nucleation and growth mechanism. For instance, the same flow rate (10 sccm) and the pressure range (a few mbar) could be responsible for the narrow size range. However, it is very difficult to address the detailed mechanism due to the complexity of the physics and chemistry in these gas mixtures.

### *3.2. Particle Size Distributions*

The general size ranges are similar for both UV and plasma experiments, but the particle size range for each particular case (temperature by metallicity) is distinct. In order to obtain more accurate size distribution, a larger area of the film (10 μm x 10 μm) was scanned for analyzing particle size, and the percentage of particles ($N/N_{total}\times100\%$) was plotted in 5 nm size bins (Figure 3). As shown in Figure 3, the haze particles are bigger and have a wider range at 300 K; the particles formed from the 100× metallicity mixture are between 35 nm and 125 nm in diameter, those from the 1000× mixture are between 60 nm and 190 nm, while those from the 10000× mixture vary from 80 nm to 130 nm. In contrast, the haze particles formed at 400 K are smaller but more uniform: 15 nm to 50 nm for the 100× mixture, 20 nm to 60 nm for the 1000× mixture, and 25 nm to 60 nm for the 10000× mixture. Compared to the 300 K and 400 K result, the particles produced at 600 K appear in the middle for both the average size and the size distribution range: 60 nm to 110 nm for the 100× mixture, 20 nm to 80 nm for the 1000× mixture, and 30 nm to 90 nm for the 10000× mixture. Unlike the plasma experiments (He et al. 2018), no bimodal size distribution is noticed in current UV experiments (Figure 3), confirming that the formed particles are mainly monomers.

We investigate a 3 (temperature) by 3 (metallicity) experimental matrix and the gas mixtures in nine experiments are compositionally different. Previous studies showed that the initial gas composition has an important impact on the particle size produced in Titan simulation experiments (Hadamcik et al. 2009, Hörst & Tolbert 2014, Sciamma-O'Brien et al. 2017). However, the results here suggest that the particle sizes are also temperature dependent. The temperature dependence of the particle size is more obvious for the particles formed in the 300 K and 400 K experiments, since the compositions do not vary much between the 300 K and 400 K experiments. Such temperature dependence was not observed in our previous plasma experiments, indicating that the temperature could play an important role in the photochemical formation of the haze particles. The temperature directly affects the energy levels, movements, and collisions of different molecules, and the vapor pressure of newly formed species, thus impacting the reaction rate coefficient, the formation and the nucleation of the particles. In the 400 K experiments, there might be more nucleation centers that induce the formation of a large number of small and uniform particles. In contrast, heterogeneous reactions on fewer nucleation centers could lead to the broader size range in 300 K and 600 K experiments, as observed in the plasma experiments (He et al. 2018). However, the reactions and the resulting compositions for each case could be totally different, so further comprehensive investigations are needed to evaluate the temperature's effect on the size distribution of the haze particles formed in different gas mixtures.

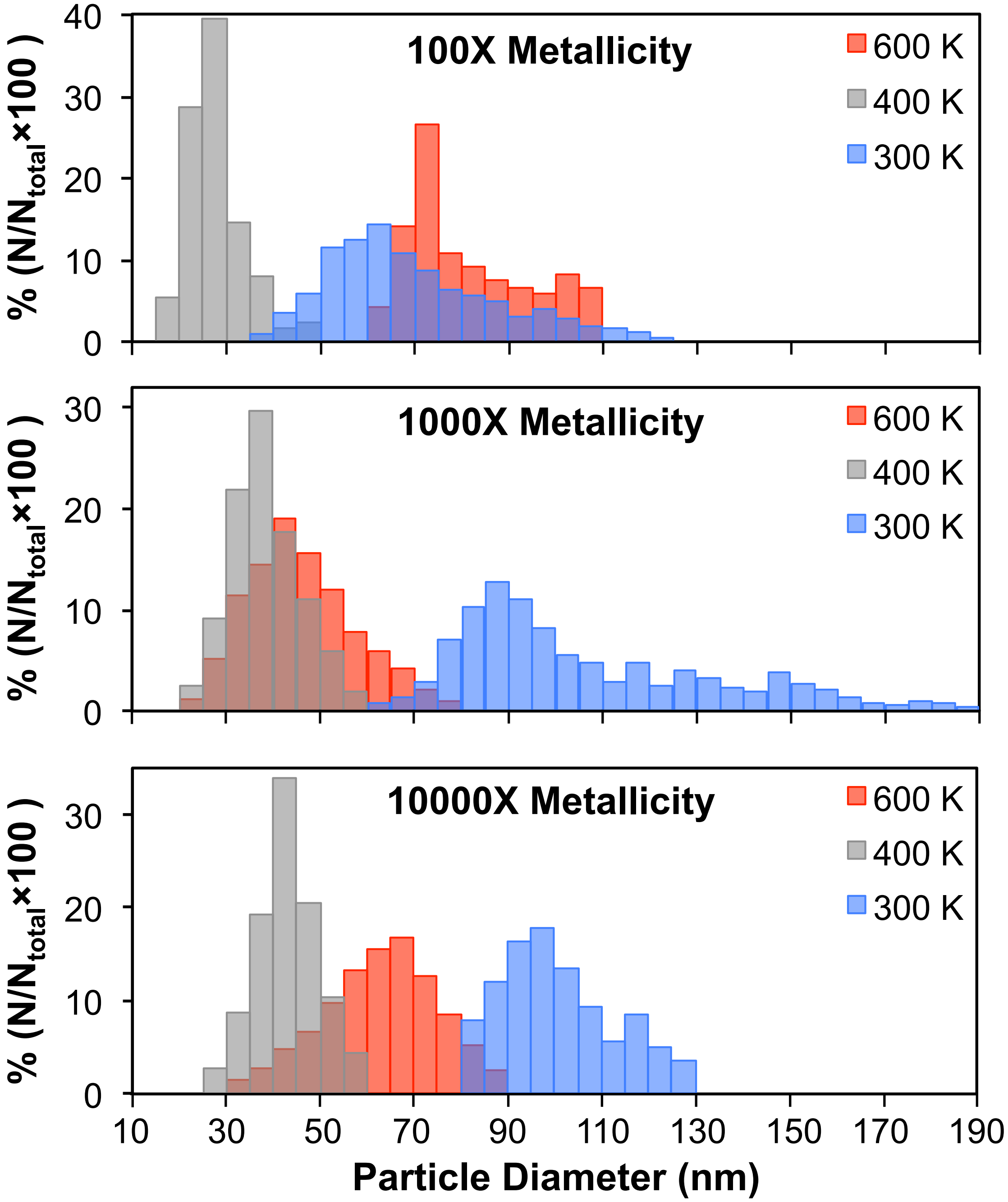

Figure 3. Haze particle size distribution in the nine UV experiments. The number of particles used for the size-distribution statistics varies from 1360 (for the 10000× experiment at 600 K) up to 64560 (for the 100× experiment at 400 K). The haze particles formed at 300 K are bigger and in wider range, while those at 400 K are smaller but more uniform; the particles produced at 600 K appear in the middle for both the average size and the size distribution range compared to those at 300 K and 400 K.

### *3.2. Haze Production Rates*

If we assume the size distribution and surface density of particles on the inside wall of the chamber are the same as that on the quartz disk, we can calculate the total volume ($V$) of the particles based on the number and size distribution we learned from the AFM images:

$$V = \sum_{i=0}^{i} \frac{\pi}{6} D_i^3 N_i \quad (1)$$

where $D_i$ is the median particle diameter in each bin, $N_i$ is the number of particles in each bin over the total available surface area within the chamber.

If we assume the particle densities are the same for all the nine cases and do not change as function of size, we can calculate the total mass of the particles by giving a density equal to that of Titan tholin sample (He et al. 2017). Previous studies show that the particle density varies with the initial gas mixture (Hörst & Tolbert 2013 & 2014, He et al. 2017). Nine gas mixtures investigated here are compositionally distinct, so the particle densities for nine experiments are unlikely to be the same. Although the constant density assumption we made here may not be correct, it allows us to estimate the production rate, and to compare to those from other experiments. The total mass ($m$) of the particles equals volume times density [$\rho$ = 1.38 g cm$^{-3}$, average density of Titan tholin samples from previous study (He et al. 2017)].

The total volume and mass of the haze particles produced in the nine UV experiments are plotted in Figure 4, and the production rates are listed in Table 1. The production rates of our previous plasma experiments (He et al. 2018, Hörst et al. 2018a) are also included in Table 1. For those plasma experiments that did not produce enough solid particles to collect and weigh, the production rates are determined by the method described above. As shown in Table 1, the production rates of the UV experiments are lower than those of the plasma experiments, except the 100× experiments at 300 K that we could not compare directly (since the production rates calculated for the 100× plasma experiment at 300 K is a lower limit). Figure 4 and Table 1 shows that the 1000× experiment at 300 K has the highest production rate (0.060 mg/hr) among the nine UV experiments. Interestingly, the 1000× plasma experiment at 300 K also has the highest production rate, although it has much higher rate (10.43 mg/hr). The 1000× experiments at 300 K have the highest haze production rate for both energy sources, indicating that we expect small, cool planets with high metallicity atmospheres to have substantial haze production. As shown in Figure 4 and Table 1, the haze production rates in the UV experiments decrease as the temperature increases for each metallicity group, except the 10000× experiments that have similar production rates at 400 K and 600 K. This temperature dependence could be related to the vapor pressure of newly-formed species. The vapor pressure of any substance increases as the temperature increases. At lower temperature (300 K), the newly-formed species have lower vapor pressure and tend to condense and/or nucleate, generating more particles; at higher temperature (400 K and 600 K), the vapor pressure of the newly-formed species increase, and these species are more likely to stay in gas phase and be removed from the system. Further compositional analysis of both the gas phase and solid phase products are required to verify this idea.

The nine experiments started from different gas mixtures, but all led to the formation of haze particles, demonstrating that there are multiple photochemical pathways for organic haze formation. For the 100× and 1000× experiments, $CH_4$ provides the carbon source for the organic haze, but the 10000× experiments have no $CH_4$ at all. For the 10000× cases, both current UV experiments and our previous plasma experiments (He et al. 2018) generate organic haze particles. The haze formation without methane in the initial gas mixture indicates that organics can be produced from other carbon species, such as CO and $CO_2$. Previous studies have shown that a variety of organic compounds can be produced in the gas mixture of $CO/N_2/H_2O$ or $CO_2/N_2/H_2O$ under UV (or plasma) irradiation (See e.g., Bar-Nun & Chang 1983, Plankensteiner et al. 2004, Cleaves et al. 2008). The result here indicates that CO and $CO_2$ could provide carbon source for photochemically produced organic hazes, and $CH_4$ is not necessarily required. It should be noted that the haze production rates are not simply a function of carbon abundance. Many factors, such as the reducing/oxidizing environment of the system, the absorption cross-sections of different reaction species, and the temperature, can affect the photochemical haze production. In addition, all 9 initial gas mixtures used here are different in composition. Therefore, further work is necessary to understand the complex photochemical processes leading to the formation of organic hazes.

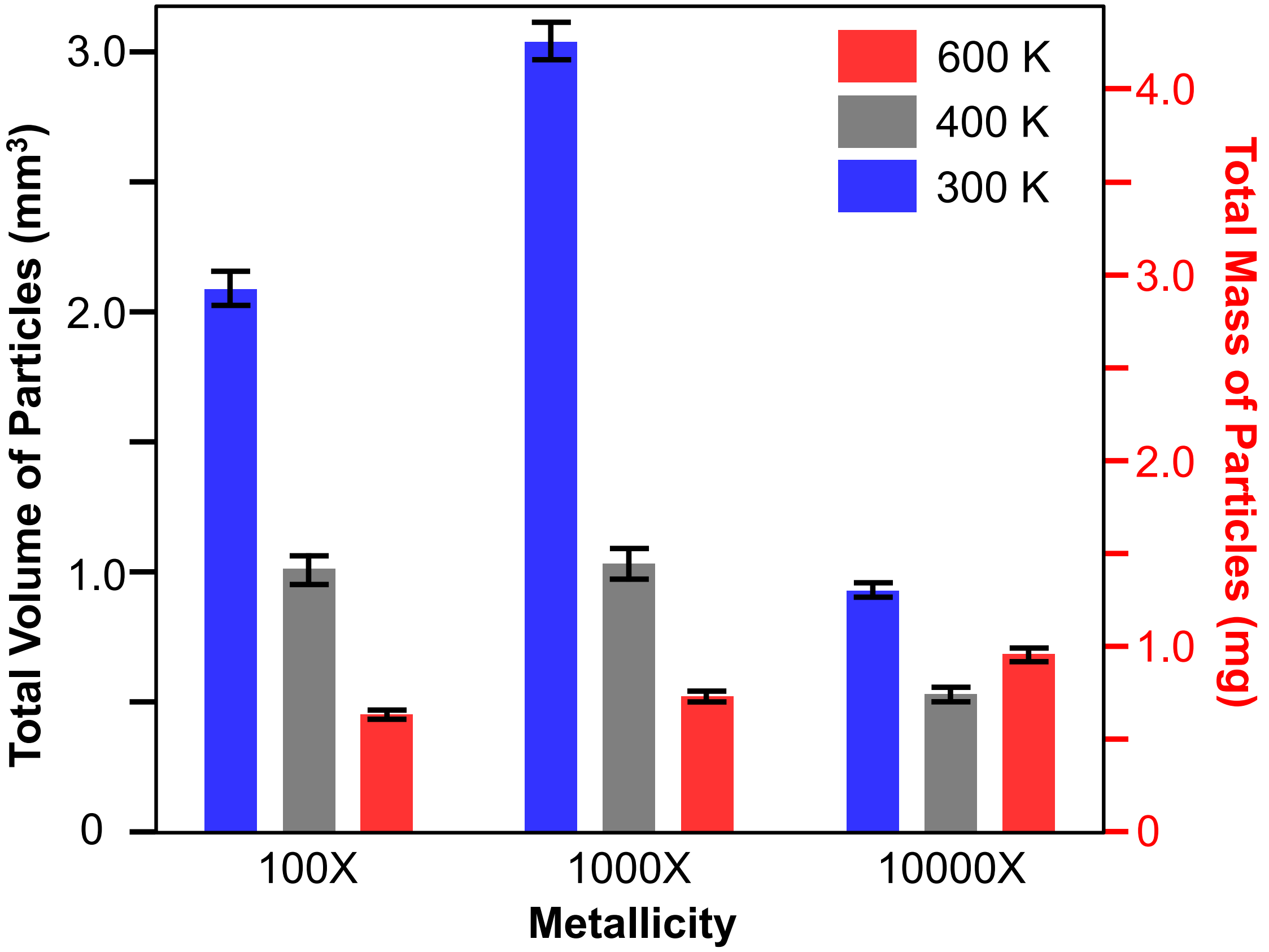

Figure 4. The total volume and the total mass of the particles produced in the nine experiments. The error bars for the total volumes are derived by propagating uncertainties of the measured particle diameters. We assume a fixed particle density ($\rho$ = 1.38 g $cm^{-3}$) to calculate total mass of the particles (the right axis). The total volume or the total mass of the particles produced in each metallicity group decrease as the temperature increases, except the 10000× experiments that have similar production rates at 400 K and 600 K. The 1000× experiment at 300 K has the highest production rate among the nine UV experiments.

Table 1. Production rates (mg $h^{-1}$) of the haze particles produced in our previous plasma experiments (Hörst et al. 2017) and current UV experiments

| | 100x | | 1000x | | 10000x | |
|---|---|---|---|---|---|---|
| | Plasma | UV | Plasma | UV | Plasma | UV |
| 600 K | 0.04 | 0.008* | 0.15 | 0.010* | 0.31 | 0.013* |
| 400 K | 0.25 | 0.019* | 10.00 | 0.021* | 0.013* | 0.010* |
| 300 K | >0.016* | 0.041* | 10.43 | 0.060* | 0.052* | 0.018* |

"*" indicates that the experiment did not produce enough solid particles to collect and weigh, and the production rate is determined by the method described in the text. The method applies to experiments with single layer of particles. ">" indicates a production rate lower limit since the AFM image (He et al. 2018) suggests there might be multiple layers of particles deposited on the substrate.

Compared to the laboratory simulations, the real atmospheres are much more complex. It is not realistic to convert the laboratory haze production rate into haze production rate in actual planetary atmospheres. However, our results here elucidate that photochemical haze production is more favorable in certain temperature and metallicity regions. Our result shows that photochemical hazes are produced in a variety of atmospheres, suggesting that haze layers could be more ubiquitous in exoplanet atmospheres than we thought. It should be noted that some minor components that are excluded in our initial gas mixtures (such as sulfur species) might be important for haze formation in exoplanet atmospheres (Zahnle et al. 2016, Gao et al. 2017). The photochemical organic haze formation in exoplanet atmospheres could affect the habitability of the planet in two aspects. First, the photochemical haze could provide the organic materials prebiotically for life to arise, like that on the early Earth. Second, haze particles can interact with light (scattering and absorption), thus influencing the energy budget and the temperature profile of the exoplanets, and potentially the habitability.

Our study also provides constraints on the particle sizes of the photochemical hazes. The haze particles from the nine experiments differ in size distribution. The haze particles of different sizes will scatter light differently, and will impact the atmospheric temperature of the exoplanets. However, very little information is known about particle sizes in exoplanet atmospheres. As we discussed in previous study (He et al. 2018), exoplanet atmospheric modelings used broad size ranges of haze particles in the models, 5 nm to 10,000 nm (see e.g., Howe & Burrows 2012, Arney et al. 2016, Gao et al. 2017). Although a variety of models have been tried to constrain particle sizes, the results are not satisfactory. For example, several studies attempted to recreate the observed transit spectrum of GJ1214b, where we see strong evidence for aerosols from the featureless spectrum, but these studies did not reach an agreement on the particle sizes: One study (Morley et al. 2015) showed that smaller particles (10 nm to 300 nm) can reproduce the featureless spectra; while the particle radii are around 500 nm from another study (Charnay et al. 2015). The size range (15 nm to 190 nm) from our current UV experiments is comparable to that (20 nm to 180 nm) from previous plasma experiments (He et al. 2018). The particles produced in both types of experiments are in the size range (10 nm to 300 nm) suggested by Morley et al. (2015), which implies that GJ1214b could have small haze particles in its atmosphere. The small particles fall in the Rayleigh scattering regime for light in visible and infrared wavelength range. In the Rayleigh scattering regime, the particles scatter short wavelength photons more efficiently and will affect the geometric albedo of the exoplanet in wavelength range from 0.4 μm to 1.0 μm (McCullough et al. 2014, Morley et al. 2015, Sing et al. 2016, Gao et al. 2017). Thus, the particles can impact the observation of the Wide-Field Infrared Survey Telescope (WFIRST), since the Coronagraph Instrument on WFIRST will directly image exoplanets in the same wavelength range (Spergel et al. 2015). Further study on their optical, thermal, and compositional properties are required to fully understand how the haze particles will affect the observations of exoplanet atmospheres.

## 4. CONCLUSIONS

We investigated the photochemical haze formation in nine different simulated exoplanet atmospheres by conducting laboratory experiments with the PHAZER chamber (He et al. 2017), and observed the particle sizes using AFM. Our result shows that photochemical hazes are produced in all nine UV experiments, and the haze production rates appear to be temperature dependent: the particles produced in each metallicity group decrease as the temperature increases. The AFM images demonstrate that the particle size (15 nm to 190 nm) varies with temperature and metallicity. For all compositions, the particles formed at 300K are the largest and those formed at 400 K are the smallest particles, despite the fact that the initial gas compositions were very similar at these two temperatures. The haze particles produced at 600 K are intermediate. The presence of haze particles significantly affects the temperature of the atmosphere and the surface of a planet can also contribute organics to the surface, thereby impacting its habitability. The result from our first experimental simulations with UV radiation provides critical inputs for exoplanet atmospheres modeling, and valuable laboratory data for future observations with facilities such as TESS, JWST, and WFIRST.

This work was supported by the NASA Exoplanets Research Program Grant NNX16AB45G. C.H. was supported by the Morton K. and Jane Blaustein Foundation.